\documentclass[12pt]{article}

\usepackage{amsmath,amsthm,amssymb}
\usepackage{epsfig}

\newcommand{\bra}[1]{\left\langle{}#1\right|}
\newcommand{\cfield}{\mathbb{C}}
\newcommand{\dff}{\sc}
\newcommand{\ee}{\mathbf{e}}

\newcommand{\hh}{\mathcal{H}}

\newcommand{\hk}{\mathcal{K}}
\newcommand{\hl}{\mathcal{L}}

\newcommand{\ket}[1]{\left| #1\right\rangle}

\newcommand{\lth}{n}

\newcommand{\ppp}[1]{{#1}{#1'}}
\newcommand{\rfield}{\mathbb{R}}
\newcommand{\rrh}{\rho}
\newcommand{\rrp}{\mathbf{p}}

\DeclareMathOperator{\trc}{Tr}

\newcommand{\cbn}{\cfield{B}_\lth}

\newcommand{\cbp}{\mathfrak{T}}
\newcommand{\bracket}[3]{\bra{#1}#2\ket{#3}}

\newcommand{\braket}[2]{\left\langle{}#1\,\right|\left.#2\right\rangle}
\newcommand{\hhp}{\mathfrak{B}}
\newcommand{\hlp}{\mathfrak{L}}
\newcommand{\ketbra}[2]{\ket{#1}\!\!\bra{#2}}

\newcommand{\mesb}{\,d\mathbf{S}_{\lth}}

\newcommand{\raypr}[1]{\ketbra{#1}{#1}}

\title{
Continuous optimal ensembles I: A geometrical characterization of
robustly separable quantum states}

\author{Rom\`an R. Zapatrin\thanks{Friedmann Lab. for Theoretical
Physics, SPb EF University, Griboyedova 30--32, 191023,
St.Petersburg, Russia; e-mail: zapatrin at rusmuseum.ru}}

\date{}

\begin{document}

\maketitle

\begin{abstract}
A geometrical characterization of robustly separable (that is,
remaining separable under sufficiently small variiations) mixed
states of a bipartite quantum system is given. It is shown that
the density matrix of any such state can be represented as a
normal vector to a hypersurface in the Euclidean space of all
self-adjoint operators in the state space of the whole system. The
expression for this hypersurface is provided.
\end{abstract}

\section{Introduction}\label{sintro}

Entanglement turned out to be a crucial resource for quantum
computation. It plays a central r\^ole in quantum communication
and quantum computation. A considerable effort is being put into
quantifying quantum entanglement.

It seems natural to focus the efforts on quantifying entanglement
itself, that is, describing the \emph{impossibility} to prepare a
state by means of LOCC (local operations and classical
communications). One may, although, go another way around and try
to quantify \emph{separability} rather than entanglement: this
turned out to be applicable for building combinatorial
entanglement patterns for multipartite quantum systems
\cite{myjmo}.

In this paper I dwell on the case of bipartite quantum systems. A
state of such system is called {\dff separable} if it can be
prepared by LOCC. In terms of density matrices that means that
$\rrp$, its density matrix, can be represented as a mixture of
pure product states. According to Carath\'eodory theorem, the
number of this states can be reduced to $\lth^4$ where $\lth$ is
the dimension of the state of a single particle.

\medskip

The idea to replace finite sums of projectors by continuous
distributions on the set of unit vectors is put forward making it
possible to provide a geometrical characterization of separable
mixed states of a bipartite quantum system. To consistently
describe the result presented in this paper recall some necessary
definitions.

\paragraph{Basics.} A density matrix $\rrp$ in the product space
$\hhp$ is called {\dff factorizable} if it is a tensor product of
density matrices, $\rrp=\rrh\otimes\rrh'$. If $\rrp$ is a convex
combination of factorizable operators, it is said to be {\dff
separable}

\begin{equation}\label{edefsepar}
 \rrp
\;=\;
\sum_\alpha\limits
\,p_\alpha\,
\rrh_\alpha\otimes\rrh'_\alpha
\end{equation}

\noindent A crucial feature of quantum mechanics, the phenomenon
of quantum entanglement, stems from the fact that there exist
density operators in the product space which are NOT separable,
they are called {\dff entangled}. A density operator $\rrp$ is
called {\dff robustly separable} if it has a neighborhood $U$ in
$\hl$ such that all operators $\rrh'\in{}U$ are separable.

\paragraph{A brief account.} In the Euclidean space $\hlp$ of
self-adjoint operators acting in the tensor product space
$\hhp=\hh\otimes\hh'$ we define a real-valued, positive functional
${\hk}:\hlp\to\rfield_{+}$ as follows

\[
{\hk}(X)
\;=\;
\iint
e^{\bracket{\ppp{\phi}}{X}{\ppp{\phi}}}
\ppp{\mesb}
\]

\noindent where the integration is taken over the torus---the
Cartesian product of unit spheres in $\hh,\hh'$, respectively, and
consider the hypersurface $\hk\subset\hlp$

\[
\hk
\;=\;
\{X\in\hlp
\,\mid\,
{\hk}(X)=1
\}
\]

\noindent Then
\begin{itemize}
  \item all robustly separable density operators in $\hh$ are in 1--1
correspondence with the points of $\hk$
  \item the density matrix associated with a point $X\in\hk$ is
the normal vector to $\hk$ at point $X$.
\end{itemize}

\section{Continuous optimal ensembles}\label{scontensemb}

To make the account self-consistent, begin with necessary
definitions. A {\dff density operator} is a non-negative
self-adjoint operator whose trace equals to 1. In particular, for
any unit vector $\ket{\phi}$ the one-dimensional projector
$\raypr{\phi}$ is a density matrix. Note that for any set of
density operators $\rrh_\alpha$ the convex combination
$\sum_\alpha{}\rrh_\alpha$ is always a density operator.

The set of all self-adjoint operators in $\hh=\cfield^\lth$ has a
natural structure of a real space $\rfield^{2\lth}$, in which the
set of all density matrices is a hypersurface, which is the zero
surface $T=0$ of the affine functional $T=\trc{}X-1$.

In this paper a geometrical characterization of separable
bipartite density operators is provided. It is based on the notion
of continuous ensembles. Generalizing the fact that any convex
combination of density operators is again a density operator, we
represent density operators as probability distributions on the
unit sphere in the state space $\hh$ of the system. Let us pass to
a more detailed account of this issue beginning with the case of a
single quantum system.

Let $\hh=\cfield^\lth$ be a $\lth$-dimensional Hermitian space,
let $\rho$ be a density matrix in $\hh$. We would like to
represent the state whose density operator is $\rrh$ by an
ensemble of pure states. We would like this ensemble to be
continuous with the probability density expressed by a function
$\mu(\phi)$ where $\phi$ ranges over all unit vectors in $\hh$.

\paragraph{Technical remark.} Pure states form a projective space
rather than the unit sphere in $\hh$. On the other hand, one may
integrate over any probabilistic space. Usually distributions of
pure states over the spectrum of observables are studied,
sometimes probability distributions on the projective spaces are
considered \cite{sqprop}. In this paper for technical reasons I
prefer to represent ensembles of pure states by measures on unit
vectors in $\hh$. I use the Umegaki measure on $\cbn$--- the
uniform measure with respect to the action of $U(\lth)$ normalized
so that $\int_{\cbn}\mesb=1$.

\subsection{Effective definition}\label{scontens}

The density operator of a continuous ensemble associated with the
measure $\mu(\phi)$ on the set $\cbn$ of unit vectors in $\hh$ is
calculated as the following (matrix) integral

\begin{equation}\label{e01integral}
  \rrh
  \;=\;
  \int_{\phi\in\cbn}\limits\;
  \mu(\phi)\,
  \raypr{\phi}
  \,\mesb
\end{equation}

\noindent where $\raypr{\phi}$ is the projector onto the vector
$\bra{\phi}$ and $\mesb$ is the above mentioned normalized
measure on $\cbn$:

\begin{equation}\label{einvarmes}
  \int_{\phi\in\cbn}\limits\;
  \,\mesb
\;=\;
1
\end{equation}

\noindent Effectively, the operator integral $\rrh$ in
\eqref{e01integral} can be calculated by its matrix elements. In
any fixed basis $\{\ket{\ee_i}\}$ in $\hh$, each its matrix
element $\rrh_{ij}=\bracket{\ee_i}{\rrh}{\ee_j}$ is the following
numerical integral:

\begin{equation}\label{e01basis}
\rrh_{ij}
\;=\;
  \bracket{\ee_i}{\rrh}{\ee_j}
  \;=\;
  \int_{\phi\in\cbn}\limits\;
  \mu(\phi)\,
  \braket{\ee_i}{\phi}
  \braket{\phi}{\ee_j}
  \,\mesb
\end{equation}

\subsection{Optimal ensembles}\label{soptens}

We need to solve the following variational problem. Given a
functional $Q$ on $L^1(\cbn)$ and given a density matrix $\rrh$ in
$\hh$, find the distribution $\mu$ on the set $\cbn$ of unit
vectors in $\hh$ such that

\begin{equation}\label{e03}
\left\lbrace
\begin{array}{l}
  \int_{\phi\in\cbn}\limits\;
  \mu(\phi)\,\raypr{\phi}\mesb
  \;=\;\rrh
   \\
   \qquad
   \\
  Q(\mu)\;\to\; \mbox{extr}
\end{array}
\right.
\end{equation}

\noindent We shall consider functionals $Q$ of the form

\begin{equation}\label{e03q}
Q(\mu)
\;=\;
\int_{\phi\in\cbn}\limits\;
q(\mu(\phi))\mesb
\end{equation}

\noindent then, according to \eqref{e01basis}, the variational
problem \eqref{e03} reads

\[
\left\lbrace
\begin{array}{l}
\int_{\phi\in\cbn}\limits\;
\mu(\phi)
  \braket{\ee_i}{\phi}
  \braket{\phi}{\ee_j}
  \mesb\;=\;\rrh_{ij}
  \\
\int_{\phi\in\cbn}\limits\;
q(\mu(\phi))\mesb
\;\to\;
\mbox{extr}
\end{array}\right.
\]

\noindent Solving this variational problem by introducing
Lagrangian multiples $X_{ij}$ we get

\begin{equation}\label{e03a}
q'(\mu(\phi))
\,-\,
\sum_{ij}
X_{ij}
  \braket{\ee_i}{\phi}
  \braket{\phi}{\ee_j}
\;=\;
0
\end{equation}

\noindent Combining the Lagrange multiples into the operator
$X=\sum_{ij} X_{ij}\ketbra{\ee_j}{\ee_i}$ turns the equation
\eqref{e03a} to \(q'(\mu(\phi))
\,=\,
\bracket{\phi}{X}{\phi}
\). Then, denoting by $f$ the inverse of $q'$ we write \eqref{e03a} as

\begin{equation}\label{e01a}
\mu(\phi)
=
  f\left(
  \bracket{\phi}{X}{\phi}
  \right)
\end{equation}

\noindent and the problem reduces to finding $\mu$ from the
condition

\begin{equation}\label{e04ini}
\int_{\phi\in\cbn}\limits\;
\mu(\phi)
\raypr{\phi}\mesb
\;=\;\rrh
\end{equation}

\noindent which according to \eqref{e01a} and \eqref{e01basis} can
be written as

\begin{equation}\label{e04}
\bracket{\ee_i}{\rrh}{\ee_j}
\;=\;
\int_{\phi\in\cbn}\limits\;
  f\left(
  \bracket{\phi}{X}{\phi}
  \right)
\raypr{\phi}\mesb
\end{equation}

\noindent It follows from \eqref{e03a} that the coefficients
$X_{ik}$ can be chosen so that $X_{ik}=\bar{X}_{ki}$. That means
that the problem of finding the optimal ensemble reduces to that
of finding the coefficients of a self-adjoint operator, that is,
to finding $\lth^2$ numbers from $\lth^2$ equations.

\subsection{Geometrical interpretation}\label{sgeominterpr}

The equation \eqref{e04} can be given a direct geometrical
meaning. Let $\hl\simeq\rfield^{\lth^2}$ be the space of all
self-adjoint operators in $\hh$. Let $f:\rfield\to\rfield$ be a
differentiable function. Consider the real valued functional
$F:\hl\to\rfield$ defined as

\begin{equation}\label{edefderiv}
  F(X)
\;=\;
\int_{\phi\in\cbn}\limits\;
  f\left(
  \bracket{\phi}{X}{\phi}
  \right)
\mesb
\end{equation}

\noindent which is well-defined as the set $\cbn$ is compact. Fix
a basis $\{\ee_k\}$ in $\hh$, then any $X\in\hl$ is defined by its
matrix elements $X_{ik}=\bracket{\ee_i}{X}{\ee_k}$, so
$\bracket{\phi}{X}{\phi}=\sum_{ik}X_{ik}\braket{\phi}{\ee_i}\braket{\ee_k}{\phi}$.
Then the expression \eqref{edefderiv} can be treated as an
integral depending on the set of parameters $\{X_{ik}\}$. We may
consider the derivatives of $F(X)$ with respect to these
variables, calculate them

\[
\frac{\partial}{\partial X_{ik}}
\,F(X)
\;=\;
\int_{\phi\in\cbn}\limits\;
\frac{\partial}{\partial X_{ik}}
\left(
\vphantom{\frac{\partial}{\partial X_{ik}}}
\,  f\left(
  \bracket{\phi}{X}{\phi}
  \right)
\right)
\mesb
\;=\;
\]
\begin{equation}\label{ederivgen}
\;=\;
\int_{\phi\in\cbn}\limits\;
\,  f'\left(
  \bracket{\phi}{X}{\phi}
  \right)
\braket{\phi}{\ee_i}
\,
\braket{\ee_k}{\phi}
\mesb
\;=\;
\end{equation}
\[
\;=\;
  \bracket{\ee_k}{
\int_{\phi\in\cbn}\limits\;
  f'\left(
  \bracket{\phi}{X}{\phi}
  \right)
\raypr{\phi}\mesb
}{\ee_i}
\]

\medskip

\noindent So, the gradient of the functional $F$ is the operator
which can be symbolically written as

\begin{equation}\label{ederivsymb}
 \nabla F
\;=\;
\int_{\phi\in\cbn}\limits\;
  f'\left(
  \bracket{\phi}{X}{\phi}
  \right)
\raypr{\phi}\mesb
\end{equation}

\noindent and effectively calculated using \eqref{ederivgen}.

\subsection{Optimal entropy ensembles}\label{soptentropens}

Let us specify the form of the optimality functional in
\eqref{e03q} assuming it to be  the differential entropy of the
appropriate distribution:

\begin{equation}\label{edefq}
q(\mu)
\;=\;
-\mu\,\ln\mu
\end{equation}

\noindent then $q'=-(1+\ln\mu)$ and we have the following $f$ for
\eqref{e04}

\[
  f(x)
\;=\;
e^{-(1+x)}
\]

\noindent Introduce, as in \eqref{edefderiv}, the functional
${\hk}:\hl\to\rfield$ on the set of all self-adjoint operators in
$\hh$ (the minus sign and the unit summand are omitted here being
a matter of renormalization):

\begin{equation}\label{edefk}
{\hk}(X)
\;=\;
\int_{\phi\in\cbn}\limits\;
\, e^{\bracket{\phi}{X}{\phi}
}
\mesb
\end{equation}

\noindent Note that $\rrh(X)=\int_{\phi\in\cbn}\limits\;
\, e^{\bracket{\phi}{X}{\phi}
}\raypr{\phi}\mesb$ is always a positive operator, then

\[
{\hk}(X)
\;=\;
\trc\int_{\phi\in\cbn}\limits\;
\, e^{\bracket{\phi}{X}{\phi}
}\raypr{\phi}\mesb
\;=\;1
\]

\noindent is a condition which defines a full-range density matrix
$\rrh(X)$ in $\hh$. On the other hand, the condition ${\hk}(X)=1$
defines a hypersurface in the Euclidean space $\hl$. Together with
the fact that $\left(e^x\right)'=e^x$ and \eqref{ederivsymb} we
come to the following

\paragraph{Statement.} Any full-range density matrix in $\rrh$ is
associated with a point on the hypersurface ${\hk}(X)=1$ and the
entries of $\rrh$ are calculated as the components of the
gradient:

\begin{equation}\label{edefrrhgrad}
  \rrh
\;=\;
\nabla{}{\hk}
\end{equation}

\subsection{The existence}\label{sexist}

Why optimal entropy ensembles do exist for all full-range density
matrices? First note that for any full-range density matrix
$\rrh=\sum{}p_k\raypr{\ee_k}$ there are infinitely many continuous
ensembles (=probability measures on $\cbn$ in our setting)
associated with it. An example of such distribution is
$\rrh=\sum\,p_k\raypr{\ee_k}=\int\mu(\phi)\raypr{\phi}\mesb$ with

\begin{equation}\label{esmeared}
\mu(\phi)
\;=\;
\frac{
\bigl((L+1)n\bigr)!
}{ L\,n!(L\,n)! }
\;
\sum_{k=1}^{\lth}\limits
\left(
p_k-
\frac{1}{L(n+1)}
\right)
\,|\braket{\ee_k}{\phi}|^{2Ln}
\end{equation}

\noindent as it follows from \cite{mygibbs}. Here $L$ is a
parameter, such that $L>\frac{1}{p_0(n+1)}$ where $p_0>0$ is the
smallest eigenvalue of $\rrh$. Any probabilistic density $\mu$
whose support is $\cbn$ is a point in the interior of the simplex
of all probabilistic measures on $\cbn$. For each probabilistic
measure on $\cbn$ its differential entropy can be calculated. The
differential entropy is, in turn, a concave function in the affine
space of probability distributions. Therefore if we have an affine
subset of of probability measure on $\cbn$, the differential
entropy takes its maximal value in the interior of the simplex of
probability measures. Now return to the condition in
\eqref{e03}---we see that it is affine. Therefore, if we know that
there exist at least one continuous ensemble representing $\rrh$
(but we know that as mentioned above), that means that there exist
a maximal entropy ensemble representing $X$, hence it has the
representation
\eqref{edefrrhgrad}.

\section{Bipartite systems}\label{sbipart}

Consider two finite-dimensional quantum systems whose state spaces
are $\hh,\hh'$. The state space of the composite system is the
tensor product $\hhp=\hh\otimes\hh'$. Denote by
$\hlp=\hl\otimes\hl$ the space of all self-adjoint operators in
$\hhp$.

\subsection{Continuous ensembles in bipartite case}\label{scontbi}

Let $\rrp$ be a robustly separable density matrix in the product
space $\hh\otimes\hh'$. Then it can be represented (in infinitely
many ways) as a continuous ensemble of pure product states.
Carrying out exactly the same reasoning as in section \ref{sexist}
we conclude that among those continuous ensembles there exists one
having the least differential entropy, this will be the ensemble
we are interested in. Like in section \ref{soptentropens},
formulate the variational problem. Let $\rrp$ be a density
operator in a tensor product space $\hhp=\hh\otimes\hh'$. The task
is to find a probability density $\mu(\ppp\phi)$ defined on the
Cartesian product $\cbp=\cbn\times\cbn$ of the unit spheres in
$\hh,\hh'$, respectively.

\begin{equation}\label{e03bi}
  \left\lbrace
\begin{array}{l}
  \int_{\ppp\phi\in\cbp}\limits\;
  \mu(\ppp\phi)\,\raypr{\ppp\phi}\ppp\mesb
  \;=\;\rrp
   \\
   \qquad
   \\
  Q(\mu)\;\to\; \mbox{extr}
\end{array}
\right.
\end{equation}

Proceeding exactly in the same way as with single particle, we get
the following representation:

\begin{equation}\label{erepbi}
  \rrp
\;=\;
  \int_{\ppp\phi\in\cbp}\limits\;
  e^{\bracket{\ppp\phi}{X}{\ppp\phi}}
\,\raypr{\ppp\phi}\ppp\mesb
\end{equation}

\noindent for some self-adjoint operator $X$ in $\hl$ whose
existence is guaranteed by the same reasons as in section
\ref{sexist}. Why such $X$ does not exist for entangled density
operators? The reason is that the set of probability distributions
among which $e^{\bracket{\ppp\phi}{X}{\ppp\phi}}$ is optimal is
simply void in the entangled case.

\subsection{Geometrical characterization of
robustly separable quantum states}\label{sgeombi}

Now we pass to the main result of this paper. Suppose we deal with
a tensor product of two Hilbert spaces $\hh,\hh'$, each of
dimension $\lth$. Consider the space $\hl$ of all self-adjoint
linear operators in the tensor product $\hhp=\hh\otimes\hh'$,
being a Euclidean space of dimension $\lth^4$. For any $X\in\hl$
we can always calculate the integral

\begin{equation}\label{edefkbi}
{\hk}(X)
\;=\;
  \int_{\ppp\phi\in\cbp}\limits\;
  e^{\bracket{\ppp\phi}{X}{\ppp\phi}}
\ppp\mesb
\end{equation}

\noindent which is always well-defined (as an integral of a
bounded function over a compact set), positive (as the exponent is
always positive) functional from $\hl$ to $\rfield_+$. Consider
the hypersurface $\hk$ in $\hl$ defined  by the equation

\[
\hk
\;=\;
\{X\in\hl
\,|\,{\hk}(X)=1
\}
\]

\noindent In any point of $\hl$ the gradient $\nabla {\hk}$ can be
calculated. In particular, at any point $X$ of $\hk$ the gradient
$\nabla {\hk}$ will be a normal vector to $\hk$. The surface $\hk$
is something given once and forever, it depends only on the
dimensionality of the state space. For any $X$ such that
${\hk}(X)=1$, we can calculate the gradient $\rrp(X)=\nabla
{\hk}\left|{}_{X}\right.$ at point $X$ Fix bases $\{\ee_i\}$,
$\{\ee'_{i'}\}$, then
$X=\sum_{\ppp{i}\,\ppp{k}}\,X_{\ppp{i}\,\ppp{k}}\ketbra{\ppp{i}}{\ppp{k}}$
and the expression \eqref{erepbi} for the operator $\rrp$ has the
following form:

\begin{equation}\label{edefgradbi}
\rrp_{\ppp{i}\,\ppp{k}}
\;=\;
\nabla {\hk}
\;=\;
\frac{\partial{{\hk}}}{\partial{X_{\ppp{i}\,\ppp{k}}}}
\end{equation}

Conversely, given a robustly separable bipartite density matrix
$\rrp$, we know that it can be represented as a convex combination
of product states: $\rrp=\sum
p_{\alpha}\rrh_{\alpha}\otimes\rrh'_{\alpha}$. Each
$\rrh_{\alpha}$ can be, in turn, represented as a non-vanishing
probability distribution \eqref{esmeared}. Then exactly the same
reasoning as in section \ref{sexist} can be carried out and there
is a point $X$ on the surface $\hk$ associated with $\rrp$. So,
together with \eqref{edefgradbi}, we have the main result:

\begin{equation}\label{emainresbi}
\bigl\{
  \mbox{robustly separable states}
\bigr\}
\quad\leftrightarrow\quad
\bigl\{
  \mbox{the points of $\hk$}
\bigr\}
\end{equation}

\section*{Summary}

A geometrical interpretation of robustly separable density
operators of a bipartite quantum system with the state space
$\hhp=\hh\otimes\hh'$ is provided. They are represented as normal
vectors to the hypersurface $\hk$ in the (Euclidean) space $\hlp$
of self-adjoint operators in $\hhp$ defined by the following
equation:

\begin{equation}\label{edefkbiconc}
\hk
\quad=\quad
\left\{
\vphantom{\int_{\ppp\phi\in\cbp}\limits}
\,X\; \right|
\left.
\;
  \int_{\ppp\phi\in\cbp}\limits\;
  e^{\bracket{\ppp\phi}{X}{\ppp\phi}}
\ppp\mesb
\;=\;1\;
\right\}
\end{equation}

\noindent where the integration is performed over the set of all
unit product vectors $\bra{\ppp\phi}\in\hhp$. Each point $X\in\hk$
is a self-adjoint operator, the parameter of the probability
distribution on the set of unit vectors which gives a density
operator $\rrp$. Furthermore, the normal vector to $\hk$ at point
$X$ is $\rrp$ itself:

\begin{equation}\label{erepbiconcl}
  \rrp
\;=\;
\nabla
{\hk}\left|{}_{X}\right.
\;=\;
  \int_{\ppp\phi\in\cbp}\limits\;
  e^{\bracket{\ppp\phi}{X}{\ppp\phi}}
\,\raypr{\ppp\phi}\ppp\mesb
\end{equation}

\paragraph{The final remark.} Given a density matrix $\rrp$ in
$\hhp$, a question arises if it is separable or not. When the
dimension of at least one of spaces $\hh,\hh'$ is 2, this question
was given an effective answer---the positive partial transpose
(PPT) criterion due to Peres-Horodecki was suggested
\cite{perehorod}. The criterion states that $\rrp$ is separable if
and only if its partial transpose $\rrp^{T_2}$ remains
non-negative matrix. In higher dimensions PPT is only a necessary
condition for a state to be factorizable as there exist entangled
density matrices whose partial transpose if positive.

Although a geometrical characterization of robustly separable
density matrices is provided, it does not solve (directly, at
least) the `inverse problem'. Nevertheless, the continuous
ensemble method presented in this paper seems to be helpful for
tackling the inverse problem as well. This issue is addressed in
the next paper on continuous ensembles.

\paragraph{Acknowledgments.} The idea to consider continuous
ensemble was inspired by the paper \cite{vidaltarrach}, where the
notion of robustness for entangled states was introduced, I am
grateful to its authors for the inspiration. Much helpful advice
from Serguei Krasnikov is highly appreciated. The financial
support for this research was provided by the research grant No.
04-06-80215a from RFFI (Russian Basic Research Foundation).
Several crucial issues related to this research were intensively
duscussed during the meeting Glafka-2004 `Iconoclastic Approaches
to Quantum Gravity' (15--18 June, 2004, Athens, Greece) supported
by QUALCO Technologies (special thanks to its organizers---Ioannis
Raptis and Orestis Tsakalotos).

\end{document}